\documentclass[%
reprint,
 amsmath,amssymb,
 aps,
]{revtex4-1}
\usepackage{epstopdf}
\usepackage{graphicx}
\usepackage{dcolumn}
\usepackage{bm}
\usepackage{xcolor}

\begin{document}


\title{Stability of coupled solitary wave in biomembranes and nerves}

\author{G. Fongang Achu}
 \affiliation{Complex Systems and Theoretical Biology Group (CoSTBiG),\\ Laboratory of Research on Advanced Materials and Nonlinear Science~(LaRAMaNS), Department
of Physics, Faculty of Science, University of Buea, PO Box 63, Buea, Cameroon}
\author{F. M. Moukam Kakmeni}%
\altaffiliation{Corresponding Author, Email:
moukamkakmeni@gmail.com}
\affiliation{Complex Systems and Theoretical Biology Group
(CoSTBiG),\\ Laboratory of Research on Advanced Materials and
Nonlinear Science~(LaRAMaNS), Department
of Physics, Faculty of Science, University of Buea, PO Box 63, Buea, Cameroon}%

\date{\today}

\begin{abstract}
\noindent

In this work, we consider the electromechanical density pulse as a coupled solitary waves represented
by a longitudinal compression wave and an out-of-plane transversal wave (i.e., perpendicular
to the membrane surface). We analyzed using, the variational approach, the characteristics of
the coupled solitary waves in the presence of damping within the framework of coupled nonlinear
Burger-Korteweg-de Vries-Benjamin-Bona-Mahony (BKdV-BBM) equation.  It is shown that, the inertia parameter
increases the stability of coupled solitary waves while the damping parameter decreases it.
Moreover, the presence of damping term induces a discontinuity of stable regions in the inertia-speed parameter
space, appearing in he form of an island of points. Bell shape and solitary-shock
like wave profiles were obtained by varying the propagation speed and their linear stability spectrum
computed. It is shown that bell shape solitary wave exhibit bound state eigenvalue spectrum,
therefore stable. On the other hand, the solitary-shock like wave profiles exhibit unbound state eigenvalue spectrum
and are therefore generally unstable.

\end{abstract}

\pacs{Valid PACS appear here}
\maketitle

\section{Introduction}

Nerve cells are encapsulated by a plasma membrane;
a thin quasi-two-dimensional layer consisting mainly of
lipids and proteins. Lipid membrane is integral parts
of every living cell; attributing an important role in signaling
integration to the lipid membrane. They have
been reported to resist pressure, stretching, tension, and
bending. It has been proposed that these perturbations,
in the form of a density or voltage pulses, may play a
major role in inter or intracellular communications \cite{F1,F2,F3} as well as nerve pulse propagations  \cite{F4,F5}.
Signal transduction
across the plasma membrane via various receptors
and ion channels has received much attention. In
a now-classic model proposed by Hodgkin and Huxley,
the ionic hypothesis is used to explain the generation
and propagation of action potential. According to the
ionic hypothesis, voltage-gated ions channels are seen as
being responsible for action potential propagation along
the plasma membrane of the neuronal axon \cite{s1, s2, s3}. The
Hodgkin-Huxley model is an electrical model based on
two conductors (cytosol and extracellular space) separated
by a capacitor (plasma membrane) with ions specific
channels (proteins). This model is the currently accepted
model, substantiated by the discovery of ion channels \cite{P5} and their crystallization \cite{P6}. However, the ionic hypothesis
has been unable to explain the non-electric phenomena
observed during the propagation of action potential: swelling of the axon, phase transition, shortening of the
axon, and adiabatic nature of action potential. 
A thermodynamic theory was therefore proposed by Heimburg
and Jackson in 2005 to address some these issues \cite{F4}. A
thermodynamic or soliton theory of nerve pulse propagation
hypothesizes that the action potential propagates in
the form of a soliton, or sound wave, along with the lipid bilayer
\cite{F1,F2,F4, F6}. According to this theory, the compression
of a lipid membrane will change its density resulting
to phase transitions from a liquid state to a gel state accompanied
by the propagation of a density pulse in the gel
state. Thus in the soliton theory, the transmission, storage,
and information processing are intrinsic properties of
the lipid bilayer. The soliton model is based on the propagation
of a localized density wave in the axonal membrane
\cite{F4, F5}. The important requirement of the model
is the empirically known lipid phase transitions slightly
below the physiological temperatures. The soliton model
has been successful in predicting the exact pulse propagation
velocities in myelinated nerves. The propagation
velocities are closely related to the lateral sound velocities
in the nerve membrane \cite{F4, F5}. Moreover, the soliton
model explains the reversible temperature and heat exchanges
observed in connection with the nerve pulse. In
the soliton theory, the appearance of a voltage pulse is
explained to be a consequence of the piezoelectric nature
of partially charged and asymmetric cell membrane
\cite{F4, F5}.\

A great effort is being devoted to the deciphering of the
mechanism of signal propagation in the axon, in particular,
the coupling between electrical and mechanical waves
in the form of the longitudinal compression waves \cite{F1, F2, F3, F4, F5, F6, F7, F8, F9,  s6, s7, F10, F11, F12}. Sound propagates well in lipid membranes in
the form of the longitudinal compression wave. What
is not is not obvious, though, is whether the out-of-plane
transversal mode can propagates in disordered systems
such as biomembrane and nerves. Many experiments
have been carried out to detect transverse wave propagation
in the lipid membrane. As early as 1988, Vogel
and M\"obius investigated surface density fluctuations in
lipid monolayers at air/water interfaces by superposing
two transverse capillary waves under resonance such that
their two wave vectors are orthogonal \cite{P1}. They observed
that the surface density fluctuations propagate
longitudinally along the lipid monolayers. In addition,
the wavelength of the transverse mode of the transverse
capillary waves was found to be modulated by the longitudinal
mode of the wave at resonance. This results
clearly indicate the inherent coupling between the longitudinal
and the transverse mode. In the soliton theory
for the action potential, a two-dimensional sound wave
propagates along the membrane plane, described by lateral
compressibility and a correlated change in density.
When applied to the axon, the action potential is considered
as a one-dimensional represented by a longitudinal compression
pulse since the diameter of the axon is assumed
to be smaller than the length of the nerve pulse \cite{F4, F8, F9,s6, s7, F10, F11, F12}. In contrast, El Hardy and Machta consider not only
the longitudinal compressional dislocations but also an
out-of-plane transversal mode (i.e., perpendicular to the
membrane surface) that Heimburg and Jackson did not
considered. This means that El Hardy and Matcha effectively
consider a three-dimensional wave propagation
along the axon. That is, two in a plane and one
out of a plane and assuming the axon to display a circular
cross-section as in the case of the soliton model, they get
rid of one dimension \cite{F13}. Thus, the electromechanical
density pulses in the nerve can effectively be considered
as a vector soliton with two components. That is,
the longitudinal component corresponding relative height
field, and the transverse components corresponding to
lateral stretch field, and have been observed during nerve
pulse propagation in garfish olfactory nerve, squid giant
axon, and hippocampal neuron \cite{F13}.\

Stimulation of peripheral nerves can be used for the
treatment of the dysfunction of the lower urinary tract,
chronic pain, epilepsy, and other neurological disorders
\cite{G6}. It is worth noting in this connection that the composition
of the phospholipid bilayer of neurons in terms of
its constitutive lipid has been strongly implicated in the
the onset of neurobiological and psychiatric disorders \cite{B}. It
will be of great interest to understand the characteristics
of signal propagation in a lipid bilayer; a ``device'' for
memory storage and information processing, as it might
play a role in the functional success of neurotherapy. As
a consequence, a right mathematical model of pulse propagation
in the lipid membrane and nerve is therefore required.
In this work, we have considered the coupled
solitary wave model for action potential derived from the
improved soliton model of nerve. We investigated in section II, using
the variational approach the characteristics of a coupled
solitary waves in the presence of damping within
the framework of coupled nonlinear Burger-Korteweg-de
Vries-Benjamin-Bona-Mahony equations (BKdV-BBM)
derived from the vector Boussinesq equation describing
the dynamics of two components nerve impulse in the soliton
model for biomembranes and nerves. In the last section,
the stability of the coupled solitary wave solution is
studied both analytically and numerically.\

\section{The coupled solitary waves model for nerve impulse transmission. } The propagation of action potential in the soliton
model for biomembranes and nerve is usually considered
as  a longitudinal compression
pulse. This is done  by assuming that the diameter of the
axon is smaller than the length of the nerve pulse. However,
recent studies show that one need not only to consider
the longitudinal compressional dislocations but also
the out-of-plane transversal mode (i.e., the mode perpendicular
to the membrane surface)
The propagation of an action potential in the soliton
model for biomembranes and nerve is usually considered
as a one-dimensional represented by a longitudinal compression
pulse; by assuming that the diameter of the
axon is smaller than the length of the nerve pulse. However,
recent studies show that one need not only to consider
the longitudinal compressional dislocations but also
the out-of-plane transversal mode (i.e., the mode perpendicular
to the membrane surface). By considering
the nerve signal \textbf{U} as an electromechanical density pulse
with two components, the improved Heimburg-Jackson
model equation is rewritten as
 \begin{eqnarray}\label{e1}
\frac{\partial^2{\textbf{U}}}{\partial{t^2}}&=&\frac{\partial}{\partial{x}}\left\lbrace\left\lgroup
c^{2}_{0}+\alpha \textbf{U}+\beta \textbf{U}\textbf{.}\textbf{U}\right\rgroup \textbf{.} \frac{\partial{\textbf{U}}}{\partial{x}}\right\rbrace
 +\nu\frac{\partial^2}{\partial{x^2}}\left\lgroup\frac{\partial{\textbf{U}}}{\partial{t}}\right\rgroup \nonumber \\ &&-\eta_{1}\frac{\partial^4{\textbf{U}}}{\partial{x^4}}+\eta_{2}\frac{\partial^4{\textbf{U}}}{\partial{x^2}\partial{t^2}}.
\end{eqnarray}

\noindent In Eq. (\ref{e1}), $x$ is the spatial coordinate along the membrane
cylinder and  $t$ is time. $\textbf{U}$ describes a two component electromechanical density pulse define as
\begin{eqnarray}\label{e2}
\textbf{U}= u(x,t)e_{r}+ v(x,t)e_{\theta},
\end{eqnarray}
where, $u(x,t)$ and $v(x,t)$ correspond to the longitudinal and  transverse components respectively. $e_{r}$ is the unit vector in the direction of $u(x,t)$ while $e_{\theta}$ is the unit vector in the direction of $v(x,t)$. $e_{r}$  and $e_{\theta}$ satisfy the orthonormal properties $e_{r}.e_{\theta}=0$, $e_{\theta}.e_{\theta}=1$, and $e_{r}.e_{r}=1$. $\nu$ is the damping coefficient while $\eta_2$  and $\eta_1$ are the  dispersion parameters.

The second dispersion coefficient $\eta_2$ is
physically related to the inertia effects of the membrane
structure and have been shown to influence the width of
the solitary pulse. In this improved model, $\eta_1$ determines the limiting velocity at high frequencies and $\eta_2$ governs how this limit is reached \cite{s6, s7} and in \cite{F12} it was established that $\eta_2$ increases the stability the nerve signals. A great deal of work has been done on the effects of damping coefficient $\nu$ on the dynamics of nerve pulse. It has been established that $\nu$ causes modulation instability in the nerve, and possible generation of localized periodic wave trains \cite{F11}. Such localized wave trains were analytically and numerically  obtained in \cite{F12}. The parameters $\alpha$ and $\beta$ describe the nonlinear
elastic properties of membranes. At a temperature slightly above
the melting transition, the lipid membrane has negative values
for the parameter $\alpha$ and positive values for the parameter $\beta$ and in low-frequency limit for  dipalmitoyl phosphatidylcholine (DPPC) membranes, the initial speed of sound $ c_0=176.6 $ m/s, $\alpha=-16.6 \frac{c_0^2}{\rho^A_0}$
, and $\beta= 79.5 \frac{c_0^2}{(\rho^A_0)^2}$
with initial density of the membrane $\rho^A_0=4.035 \times 10^{-3}$  $gm^{-2}$.\

If we assume a small oscillation of the density pulse, $\textbf{U}\rightarrow \epsilon \textbf{U}$ and the system parameters $c_o\rightarrow \epsilon c_0$, $\eta_2\rightarrow \epsilon^{-2} \eta_2$, $\nu\rightarrow \epsilon\nu$, and the transformation
  \begin{equation}\label{e3}
  y=\epsilon(x-t), \hskip 0.25truecm s=\epsilon^3t,
 \end{equation}
 where $\epsilon$ and $\epsilon^3$ are chosen in such a way to
 balance the effects of nonlinearity and damping, and substituting into Eq. (\ref{e1}), terms of order $\epsilon^4$ give
\begin{widetext}
 \begin{equation}\label{e4a}
 \frac{\partial{u}}{\partial{s}}+\frac{1}{2}(c_0^2 +
 \beta u^2)\frac{\partial u}{\partial y}-\frac{\eta_1}{2}\frac{\partial^3u}{\partial
 y^3}- \eta_2\frac{\partial^3u}{\partial
 y^2\partial s} -\frac{\mu}{2}\frac{\partial^2u}{\partial{y}^2}+\frac{\beta}{2}v^2\frac{\partial u}{\partial y}=0,
 \end{equation}
 \begin{equation}\label{e4b}
 \frac{\partial{v}}{\partial{s}}+\frac{1}{2}(c_0^2 +
 \beta v^2)\frac{\partial v}{\partial y}-\frac{\eta_1}{2}\frac{\partial^3v}{\partial
 y^3}- \eta_2\frac{\partial^3v}{\partial
 y^2\partial s}-\frac{\mu}{2}\frac{\partial^2v}{\partial{y}^2}+\frac{\beta}{2}u^2\frac{\partial v}{\partial y}=0.
 \end{equation}
\end{widetext}

 The nonlinear coupled Burger-KdV-Benjamin-Bona-
 Mahony equations  (\ref{e4a}) and (\ref{e4b})  describe the dynamics
 of two coupled solitary waves in the improved soliton
 model of nerves. It can be used to model the dynamics of
 the mechanical fields (i.e., relative height field and the
 lateral stretch field) that has been observed in a neuron.
 It is instructive to note that, a linearly coupled KdV
 equations have been used to model two layer settings in
 various physical systems, such as stratified fluids with
 a superimposed shear flow, and in dual-core optical
 waveguides carrying ultrashort pulses. In such a system,
 two types of vector soliton solutions (i.e., symmetric
 and asymmetric solitons) are usually obtained using
 the variation approximation method \cite{S14}. In addition,
 the coupled KdV-Benjamin-Bona-Mahony equation
 have been used to model the motion of small-amplitude
 long waves on the surface of an ideal fluid under the
 gravity force, and in situations where the motion is
 sensibly two dimensional. In such a system, one of the
 component models the elevation of the fluid surface from
 the equilibrium position while the other component
 models the horizontal velocity in the flow \cite{G14}.\\

\section{Variational formalism and analytical solution to the coupled model equation}

The understanding of physical, and, in particular, nonlinear,
phenomena through conservation laws and variational
formalism is probably the most fundamental and
universal approaches to theoretical physics \cite{k1}. In this
context the natural question is what kind of information
can we potentially get from knowing the energy carried
by the solitary waves in the nerve fiber?. This approach
is widely acknowledged in such branches of nonlinear
science as fluid dynamics \cite{k1, k2}, plasma physics
\cite{k3, k4}, and in the recent wave of activity on the dynamics
of trapped Bose condensates, as well as in the
the general context of the Hamiltonian systems \cite{k5}. Here
however, we are interested in the solitary wave solution
of the coupled system (\ref{e4a}) and (\ref{e4b}) via variational
method. To proceed we first  applied traveling wave
ansatz method and obtained a set of coupled modified
Lienard's equations. It is worth noting that, the modified
coupled Lienard’s equation has been used to model
the dynamics of two coupled bursting neurons by
considering a set coupled Hindmarsh-Rose (HR) neuron
model subjected to an external periodic excitation. Numerical
simulations revealed the existence of some bifurcation
structures including saddle-nodes, symmetry breaking
and period-doubling route to chaos \cite{k6}. It is
imperative to note that the addition of the frictional term
to the classical soliton model gives an additional term to
the Lagrangian of the system; the so-called frictional potential
term. In order to obtain the Lagrangian density
with the damping term, the conventional Lagrangian formalism
is inconsistent. Therefore we employed the differential
approach which consists of multiplying the conventional
Lagrangian with an exponential factor \cite{S14, F15, F16, F17, F18, F19}.\

Proceeding as described above, we first seek travelling wave solution of the coupled system (\ref{e4a}) and (\ref{e4b}) by using the traveling-wave ansatz, $u=u(y-\xi s)=u(z)$ and $v=v(y-\xi s)=v(z)$ (where $\xi$ is the solitary wave velocity). Substituting into Eqs. (\ref{e4a}) and (\ref{e4b}), we obtained the coupled Lienard's equation

\begin{equation}\label{e6a}
\frac{\partial^2{u}}{\partial{z^2}}+\frac{\gamma}{\gamma_o}\frac{\partial u}{\partial z}+\frac{c_o^2-2\xi}{2\gamma_o}u+\frac{\beta}{6\gamma_o}u^3+\frac{\beta}{2\gamma_o}uv^2=0,\\
\end{equation}\
\begin{equation}\label{e6b}
\frac{\partial^2{v}}{\partial{z^2}}+\frac{\gamma}{\gamma_o}\frac{\partial v}{\partial z}+\frac{c_o^2-2\xi}{2\gamma_o}v+\frac{\beta}{6\gamma_o}v^3+\frac{\beta}{2\gamma_o}v u^2=0,
\end{equation}
where $\gamma=\frac{-\nu}{2}$ and $\gamma_o=\frac{2\xi\eta_2-\eta_1}{2}$. In the variational Lagrangian approach, Eqs. (\ref{e6a}) and (\ref{e6b})  are restated as a variational problem in terms of the Lagrangian $L^{'}$ given by
\begin{eqnarray}\label{e8}
L^{'}&=&\left \lgroup \frac{1}{2}\left \lbrace u_z^2+v_z^2 \right\rbrace -\frac{c_o^2-2\xi}{4\gamma_o}\left\lbrace u^2+ v^2 \right \rbrace\right \rgroup e^{\frac{\gamma}{\gamma_o}z} \nonumber \\&&-\left \lgroup \frac{\beta}{24\gamma_o}\left\lbrace u^4+ v^4 \right\rbrace+\frac{\beta}{2\gamma_o}u^2v^2 \right \rgroup e^{\frac{\gamma}{\gamma_o}z}.
\end{eqnarray}
In Eq. (\ref{e8}), the exponential  term $ e^{\frac{\gamma}{\gamma_o}z}$ denotes the damping characteristics of the coupled solitary waves. In order to analyze the dynamics of the coupled solitary waves, we use a reduced variational principle and assume solitary  wave ansatz  as a trial function. Thus we choose an ansatz of the form
\begin{equation}\label{e9}
(u, v )=( A, B) \textmd{sech} (az),
\end{equation}
and substituting this into  Eq. (\ref{e8}), we obtain the effective Lagrangian $L$, given by
\begin{subequations}
\begin{eqnarray}\label{e10}
L&=&\int^{\infty}_{-\infty}L^{'} dz, \\
&=&\left \lgroup \frac{3\gamma-2}{3\gamma}+\frac{3\gamma^2_o(2\xi-c_o^2)-\gamma^2}{6\gamma\gamma_o^2a^2}\right \rgroup ( A^2+B^2)  \nonumber \\&& \times \quad \frac{\gamma \pi}{2\gamma_o} \textmd{cosec}\left \lbrace \frac{\gamma \pi}{2\gamma_o a}\right \rbrace  \nonumber \\&& -\frac{\beta}{72\gamma_o^3}\left\lgroup \frac{\gamma^2}{a^4}-\frac{4\gamma^2_o}{a^2}\right \rgroup (A^4 +B^4+12A^2B^2)  \nonumber \\&& \times \quad \frac{\gamma \pi}{2\gamma_o} \textmd{cosec}\left \lbrace \frac{\gamma \pi}{2\gamma_o a}\right \rbrace.
\end{eqnarray}
\end{subequations}

It should be noted that, the values of the amplitudes $A$ and $B$, and the inverse width $a$,  for the soliton solutions are real, and can be gotten in terms of the system parameters by using the Euler-Lagrangian equations $\frac{\partial L}{\partial A}=\frac{\partial L}{\partial B}=\frac{\partial L}{\partial a}=0$. Thus, we obtain the system of equations
\begin{widetext}
\begin{subequations}
\begin{eqnarray}
\frac{3\gamma-2}{3\gamma}+\frac{3\gamma_0^2(2\xi-c_0^2)-\gamma^2}{6\gamma\gamma_0^2a^2} -\frac{\beta}{36\gamma_0^3}\left\lbrace \frac{\gamma^2}{a^4}-\frac{4\gamma^2_o}{a^2}\right \rbrace \left\lbrace A^2+6B^2\right \rbrace =0 \label{e12},\label{e13a}\\
\frac{3\gamma-2}{3\gamma}+\frac{3\gamma_0^2(2\xi-c_0^2)-\gamma^2}{6\gamma\gamma_0^2a^2}-\frac{\beta}{36\gamma_0^3}\left\lbrace \frac{\gamma^2}{a^4}-\frac{4\gamma^2_o}{a^2}\right \rbrace \left\lbrace B^2+6A^2\right \rbrace=0, \label{e13b}\\
\left\lgroup\frac{3\gamma-2}{3\gamma}+\frac{\gamma^2-3\gamma_0^2(2\xi-c_0^2)}{6\gamma\gamma_0^2a^2}\right \rgroup \left \lbrace A^2+B^2 \right \rbrace -\frac{\beta}{72\gamma_0^3}\left\lbrace \frac{3\gamma^2}{a^4}-\frac{4\gamma^2_o}{a^2}\right \rbrace \left\lbrace A^4+ B^4+12A^2B^2\right \rbrace=0. \label{e13c} 
\end{eqnarray}
\end{subequations}
\end{widetext}
\
Note that, in order to obtain  Eq. (\ref{e13c}), we  assumed a small oscillation in the inverse pulse width $a$, such that $\textmd{cot}\left \lbrace \frac{\gamma \pi}{2\gamma_0 a}\right \rbrace \approx \frac{2\gamma_0 a}{\gamma \pi}$. Solving simultaneously for $A$ and $B$ in Eqs.  (\ref{e13a}) and  (\ref{e13b}) we obtain 
\begin{eqnarray}\label{e15}
(A, B)=\pm\sqrt{\frac{6\gamma_o a^2 (2\gamma_o^2a^2\lbrace 3\gamma -2\rbrace+ 3\gamma_o \lbrace 2\xi-c_o^2\rbrace)}{7\beta\gamma \left \lbrace \gamma^2-4\gamma_o^2a^2\right \rbrace}}.\qquad
\end{eqnarray}
Substituting Eqs. (\ref{e15}) into Eq. (\ref{e13c}), and solving the resulting equation for $a^2$, we obtain two roots given by
\begin{subequations}
\begin{eqnarray}
a^2=-a_1+\sqrt{\frac{a_1^2-4a_2a_0}{a_0}},\label{e16} \\
a^2=-a_1-\sqrt{\frac{a_1^2-4a_2a_0}{a_0}}, \label{e17}
\end{eqnarray}
\end{subequations}

where
\begin{subequations}
\begin{eqnarray}\label{e18}
a_o=-8\gamma_o^4(3\gamma-2), \quad\\
a_1=-2\gamma_o^2\gamma^2(3\gamma-2)+24\gamma_o^4(2\xi-\gamma^2_o)-4\gamma^2,\\ \quad and \quad
a_2=4\gamma^4-15\gamma^2_o\gamma^2(2\xi-c_o^2).
\end{eqnarray}
\end{subequations}

Note that the quantity $a_1>0$, and that $\sqrt{\frac{a_1^2-4a_2a_0}{a_0}}<< a_{1}$. As a consequence, the two roots of $a^{2}$ are approximately equal.
\begin{figure}[ht]
\includegraphics[width=3.5in, height= 3.5in]{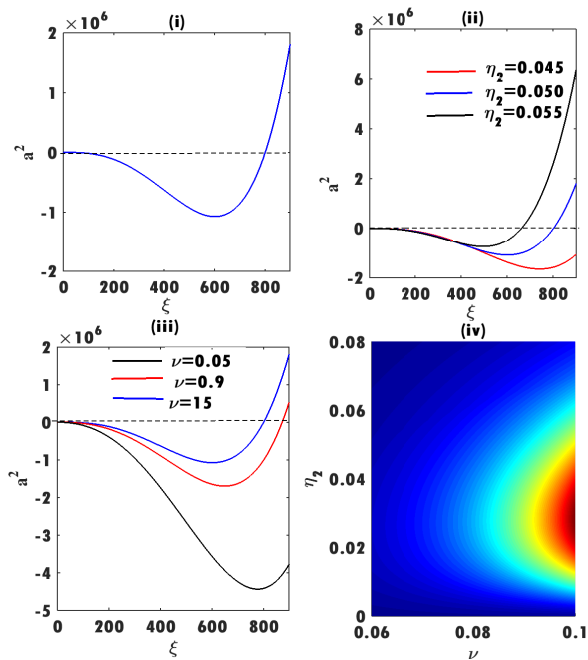}
\caption{\label{fig1}Evolution of the square of the soliton width as a function of the propagation speed $\xi$, as predicted by Eq. (\ref{e17}). (i) The order parameters are $\nu=0.05$, $\eta_1=0.05$, $\eta_2=0.05$. For $a^2<0$, that is complex pulse width (which has no physical significance), Eq. (\ref{e9}) describe  periodic solutions for real amplitudes $A$ and $B$. While for $a^2>0$, that is real pulse width Eq. (\ref{e9})  describe pulse solitons.  (ii) The effect of inertia parameter on the dynamics of the soliton. Increasing the inertia parameter leads to an increase in the stability of the soliton pulse. (ii) The effect of damping parameter on the dynamics of the soliton pulse.(iv) The level-zero contour plot of Eq. (\ref{e16}). The other model parameters
areare $\eta_1=0.05$, $c_0=176.6$,
 $\rho^A_0=4.035 \times 10^{-3}$, $\beta= 79.5 \frac{c_0^2}{(\rho^A_0)^2}$. The level-zero contour  depicts all the solution to the equation $a^2_0=F(\eta_2, \nu)$ in the contemplated parameter range. Increasing the damping parameter leads to a decrease in the stability of the soliton pulse.}
\end{figure}

Now let turn to the  analysis of the amplitudes  $(A, B)$  of the soliton as depicted by Eq. (\ref{e15}). Since $(A, B)$  and $a^2$ must be real for a pulse soliton, then from Eq. (\ref{e15}) , it easy to deduce that $a^2$ is bounded  within the region $\frac{\nu^2}{4(2\eta_2\xi-\eta_1)}< a^2<\frac{c_o^2-6\xi}{4+3\nu}$. Since $\nu >0$, it follows that the propagation speed $\xi$ is also bounded in the region $\frac{\eta_1}{2\eta_2}<\xi<\frac{c_o^2}{6}$. Figure 1 shows the evolution $a^2$ against $\xi$. In Fig. 1(i), solitary pulse  are stable for   $\xi>800$, and unstable for  $\xi<800$ for the given membrane parameters. On the other hand, Fig.1 (ii) and Fig. 1(iii) shows the effect of the inertia parameter $\eta_2$ and damping parameter $\nu$  on the stability of the  soliton pulse. The plot clearly shows that increasing $\eta_2$ increases the stability of the  soliton pulse while increasing $\nu$ decreases its stability. As a consequence, stable soliton pulse solutions describe by Eq. (\ref{e9}) is possible if there is a balance between damping and inertia effects of the lipid membrane.\\
 
 In order to further understand the dynamics of the solutions with respect to the pulse width, we turn our attention to the parameter space plot. First, we find the bifurcation points by setting $a^2(\eta_2, \xi, \nu)=0$ and we obtain the fixed points in terms of the parameters $\eta_2$ and $\nu$ by solving for $\xi$. We then plot the $3D$ surface $a_0^2=F(\eta_2, \nu)$ as a function of two variables $\eta_2$ and $\nu$ for the bifurcation point $\xi=\xi_0$. Figure 2(iv) shows the 3D surface $a_0^2=F(\eta_2, \nu)$ as introduced in Eq. (\ref{e16}). This surface depends on two parameters $\eta_2$ and $\nu$. The actual colours of any spot $(\eta_2, \nu, a_0^2)$ $ \epsilon$ $\Re^3$ on the surface depends on the height or level of $a_0^2$ above the $\eta_2-\nu$ plane. In this particular graph, the ``height'' of $a_0^2$ ranges from $0-10000$ in the surface plot. \

So far we have analyzed the solutions obtained using
the square of the pulse width described by Eq. (\ref{e16}). If we considered
that the amplitudes $A$ and $B$ are real functions and that $a^2>0$, then Eq.(\ref{e9}) is said to describe solitary wave solution which can be symmetric (i.e., $A =B$) or antisymmetric (i.e., $A=-B$ ) \cite{S14}. Hence, from  Eqs. (\ref{e15}) and Eqs.(\ref{e9}), the two-component soliton model admits three possible types of solutions: dark-dark, bright-bright,  and bright-dark solitary wave solutions. The bright-dark solitary wave solution correspond to the antisymmetric case while the dark-dark, and bright-bright solitary wave solutions corresponds to the symmetric case. It should noted here that the bright-dark mode has been recorded experimentally in the garfish olfactory nerve, squid giant axon, and hippocampal neuron \cite{F1, F2, F3, F4, F13}. 

\begin{figure}[ht]
 \includegraphics[width=3.5in, height= 3.2in]{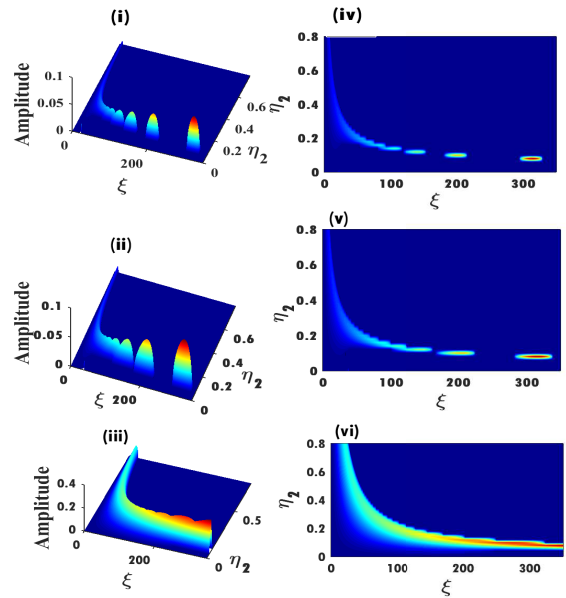}
 \caption{\label{fig11} The evolution of soliton amplitude with propagation speed $\xi$ and initial parameter $\eta_2$ for different values of the damping coefficient $\nu$ and their corresponding contour plot in the  $\eta_2-\xi$ space of Eq. (\ref{e15}): (i)-(iv) $\nu=0.05$, (ii)-(v) $\nu=0.08$ and (iii)-(vi) $\nu=0.5$. The other model parameters
 are
 $\rho^A_0=4.035 \times 10^{-3}$, $\beta= 79.5 \frac{c_0^2}{(\rho^A_0)^2}$. The amplitude response as a function of the propagation speed and the initial parameter shown the threshold and saturation effect. The evolution of amplitude as a function of damping parameter $\nu$ show the effect of discontinuity in the propagation speed.}
 \end{figure}

 \begin{figure}[ht]
  \includegraphics[width=3.5in, height= 3.5in]{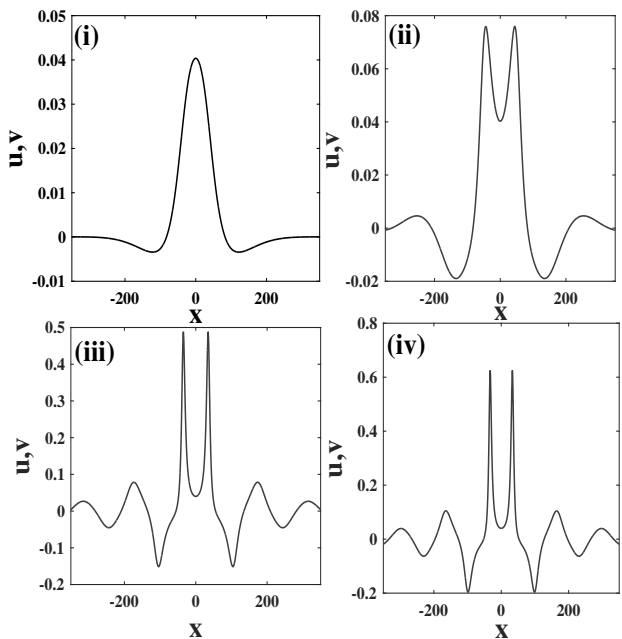}
  \caption{\label{fig2} The evolution of pulse shape as a function of  propagation speed $\xi$ within one of the  one of stable island describe in figure 4(iv): (i) $\xi=165.8$, (ii) $\xi=165.5$,(iii) $\xi=165.0$ and $\xi=164.9$. The other model parameters
  are $\epsilon=0.001$, $t=0$, $\eta_1=0.045$, $\nu=0.05$, 
   $\rho^A_0=4.035 \times 10^{-3}$ and $\beta= 79.5 \frac{c_0^2}{(\rho^A_0)^2}$. The solitary wave evolve into a solitary oscillatory shock-like waves as the propagation speed is gradually reduced. Waves with smaller propagation speed turns to propagates with higher amplitudes.}
  \end{figure}

Now consider the behavior of the soliton amplitudes
with the parameters of the system. Analysis of the soliton
amplitudes can be used to predict nonlinear phenomenon
such as symmetry breaking or discontinuity. It should
be noted that in \cite{S14}, symmetry breaking in a linearly
coupled Korteweg-de Vries system was observed. It was
shown that the linear KdV equation exhibit symmetry
breaking depending on the propagation velocity. Also in
Ref. \cite{S15}, the authors investigated numerically the existence,
and stability of solitons in parity-time-symmetric
optical media characterized by a generic complex hyperbolic
refractive index distribution and fourth-order 
diffraction. They showed that the fourth-order diffraction
coefficient greatly alters parity-time-breaking points.
In lipid membrane, discontinuity or non-linearity in state
diagram has been studied both theoretically and experimentally \cite{F1, F2, F3, F4, F5, S16}, and it is thought to results from the
liquid-expanded-liquid condensed phase change in lipid
monolayers. In fact, it is reported that the velocity of
sound decreases discontinuity as the system goes from LE
phase to LE-LC coexistence phase in liquid monolayers.
Figures 3 (iv) and 3 (v) exhibits in parameter space $(\eta_2, \xi)$
the stable and unstable regions. One observes the discontinuity
of the stable region, which appear like the island
of points when the damping parameter is varied. It should
be noted that discontinuity in speed implies a discontinuity
in the compressibility. Discontinuity in biomembranes
indicates an abrupt-phase transition, and near
phase transitions, stable solitary waves can propagate
along the nerve fiber. In order to investigate stable solitary
waves near the region of discontinuity, we plotted the
soliton profile within the stable region in Fig.3 (iv) for
different values of the propagation speed (see Fig.3). It is
worth noting that as the speed of the wave is gradually reduce,
the single pulse created evolved into an oscillatory
solitary shock-like wave. Figure 2 (vi), exhibits continuity
in the stable region within the parameter space
$(\eta_2, \xi)$ resulting from an increase in the damping coefficient.
In Refs. \cite{S16, S17}, Shrivastava \emph{et al.} have recently
found that a two-dimensional solitary shock wave reminiscent of propagating spikes in nerves can be induced in
monolayer lipid membrane near a phase transition. These
waves have a threshold for excitation and an upper bound
on the maximum amplitude (all-or-none) and are self-sustaining.
In \cite{S17}, Shrivastava further proposed that
the observed solitary shock waves at lipid interface also
provide evidence for the detonation of shock waves at
such interfaces. That is, the self-sustaining shock wave
utilizes the latent heat of phase transition of the lipids
(i.e., chemical energy stored in lipids interface) reinforcing
it in the process. One needs to stress here that the general form of the nonlinear evolution we have considered
allows for other possible sources of chemical energy
that can reinforce the propagating shock wave. Furthermore,
shock waves have been reported to be signature
for traumatic brain injury \cite{SS1}. It is well known that
exposure of biological cells to shock waves causes damage
to the cell membrane, however, the mechanisms by
which damage is caused and how it depends on physical
parameters of the shock waves such as shock waves velocity,
shock waves duration, shock waves shape, and amplitudes
is poorly understood. Figure 5 shows generation
of solitary shock-like waves at particular propagation velocity,
with different amplitudes and shock waves profiles.
Sliozberg \emph{et al.}  demonstrated numerically using coarse-grained
model of lipids vesicle the principle of damage
induced by shock waves by direct passage through the
cranium. The results show that the structural integrity
of the lipid vesicles is altered as pores are formed in the
lipid membrane. As a result, the membrane becomes
permeable to sodium, potassium and calcium ions and
therefore a possible source of chemical energy for the
shock waves \cite{SS1}. It should be stressed that, a number
of scientists have shown that membranes can display ion channel-
like current traces even in the complete absence
of proteins provided only that the membrane is close to a
melting transition \cite{S18, S19, S20}. These events are called ``lipid
channels'' and consists of small pores or defects in the
lipid membranes. We proposed that the observed solitary
shock-like wave in the coupled-model in might be
responsible for the defects in the lipid membrane near
phase transitions.

Consider the interaction of the two coupled solitary waves. From Eqs. (\ref{e8}), (\ref{e9}), and (\ref{e10}), and noting that $A^{2}=B^{2} $, we obtain the  effective potential energy function  $U_{s}$ of the coupled solitary waves
\begin{eqnarray}
U_s&=& \frac{\gamma^2_o\gamma \pi(c_o^2-2\xi)+\gamma^3 \pi}{2\gamma\gamma_o^3a^2}\textrm{cosec}\left \lbrace \frac{\gamma \pi}{2\gamma_o a}\right \rbrace A^{2} \nonumber \\&& +\frac{\beta \pi\gamma}{72\gamma_o^4}\left\lbrace \frac{\gamma^2}{a^4}-\frac{4\gamma^2_o}{a^2}\right \rbrace     \textrm{cosec}\left \lbrace \frac{\gamma \pi}{2\gamma_o a}\right \rbrace A^{4}.
\end{eqnarray}

\begin{figure}[ht]
  \includegraphics[width=3.5in, height= 2.5in]{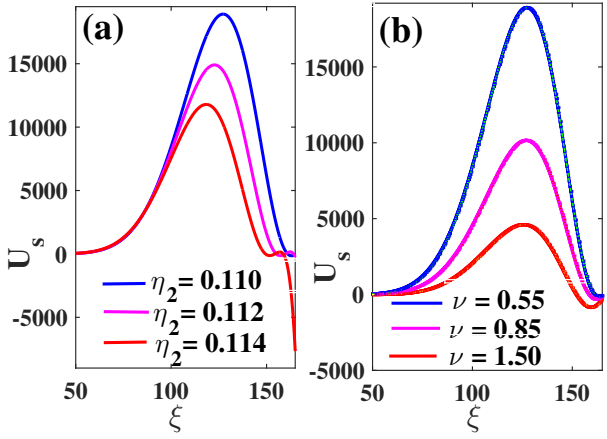}
  \caption{\label{fig6} The energy of the two solitary waves $U_s$ as a function of propagation speed $\xi$. (a) The  effect of the inertia parameter $\eta_2$  for $\nu=0.55$. (b) The effect of the damping parameter $\nu$ for $\eta_2=0.110$. The energy of the two solitary waves decrease as the damping coefficient and inertia coefficient increases. The other model parameters are  $\eta_1=0.045$, $c_0=176.6$, $\rho^A_0=4.035 \times 10^{-3}$ and $\beta= 79.5 \frac{c_0^2}{(\rho^A_0)^2}$. }
  \end{figure}

In order to conveniently analyze the characteristic of energy function, we plot the curve of $U_{s}$ against $\xi$. In view of Fig. \ref{fig6}, $U_s$ is positive in modeling region and therefore the potential function of interaction is always negative. Thus, the interaction between the two solitary waves is attractive; this agrees with the fact of neural coupling oscillation. \

In this section, we analyzed the characteristics of two
coupled solitary waves in nerve impulse transmission. We
employ an approximate approach using the variational principle
and obtained a solution for the coupled solitary
wave within the framework of the soliton model described
by coupled KdV-BBM equations in the presence of damping.
We showed that, a small friction potential present
in the coupled solitary wave model induced discontinuity
(or phase transitions) in the amplitude (density change)-
velocity parameter space, a reminiscent for propagation
of solitary shock-waves and a single soliton pulse. The relevant biological implication is also discussed. In the
next section we will study linear stability analysis of
the solitary waves solution obtained in this section.

\section{Linear Stability Analysis }
In non-linear waves, solitons and their linear stability
properties are crucial. In integrable systems, solitons
admit analytical expressions and they are generally stable
against small perturbations. In non-integrable systems,
solitons generally do not admit analytical expressions
and can be either stable or unstable. The understanding
of many properties of these nonlinear waves
can be achieved through the calculation of their spectra. Indeed,
spectral analysis of the equations appearing after
linearization of the governing non-linear equations near,
the solitons has become widespread tools. Bound eigenstates,
which do not grow with propagation or in time,
of the linear problem, are often called internal modes. Internal
modes supported by the single mode soliton in the
soliton model of nerve has been studied in \cite{F12}. In this
section, we compute the linear-stability spectrum of the solitary
wave obtained; which consists of eigenvalues of the
linear-stability operator of the solitary wave \cite{F20}. The
linear-stability spectrum contains valuable information
on the solitary wave. For instance, if this spectrum contains
eigenvalues with positive real parts, then the solitary
wave is linearly unstable. In this case, the largest
real part of the eigenvalues gives the maximal growth
rate of perturbations. If the spectrum contains purely
imaginary discrete eigenvalues, these eigenvalues are the
internal modes \cite{F20, F21}. In particular, persistent oscillations
of the soliton width and position are due to these
modes.

To proceed, we consider a small perturbation in $u$ and $v$ given by $u=u_0+\varphi u_1$ and $v=v_0+\varphi v_1$
where $\varphi <<1$ and $u_0$ and $v_0$ is the solution to the Eqs. (\ref{e4a}) and (\ref{e4b})  given by Eq. (\ref{e9}). Substituting $u=u_0+\varphi u_1$ and $v=v_0+\varphi v_1$ into Eqs. (\ref{e4a}) and (\ref{e4b})
and linearising about $\varphi$ the linearise system
\begin{widetext}
\begin{subequations}
\begin{eqnarray}\label{g}
\frac{\partial u_1}{\partial s}+\frac{1}{2}\left \lbrace c_0^2+\beta(u_0^2+v_0^2) \right\rbrace \frac{\partial u_1}{\partial y}+\beta(u_0u_1+v_0v_1)\frac{\partial u_0}{\partial y}-\frac{\eta_1}{2}\frac{\partial^3 u_1}{\partial y^3}-\eta_2\frac{\partial^3u_1}{\partial y^2\partial s}-\frac{\nu}{2}\frac{\partial ^2 u_1}{\partial y^2}=0,\label{f1}\\
\frac{\partial v_1}{\partial s}+\frac{1}{2}\left \lbrace c_0^2+\beta(u_0^2+v_0^2) \right\rbrace \frac{\partial v_1}{\partial y}+\beta(u_0u_1+v_0v_1)\frac{\partial v_0}{\partial y}-\frac{\eta_1}{2}\frac{\partial^3 v_1}{\partial y^3}-\eta_2\frac{\partial^3v_1}{\partial y^2\partial s}-\frac{\nu}{2}\frac{\partial ^2 v_1}{\partial y^2}=0.\label{f2}
\end{eqnarray}
\end{subequations}
\end{widetext}
Since the  linearisation is near a solitary wave, i.e, Eq. (\ref{e9}), then Eqs. (\ref{f1}) and (\ref{f2}), reduces into an eigenvalue problem: 
\begin{equation} \label{f3}
Q\textbf{V}=\lambda \textbf{V}.
\end{equation}
In Eq. (\ref{f3}), $\textbf{V(\textbf{y})}=(u_1, v_1)^T $ is the eigenfunction, $\textbf{y}=(y_1, ..., y_N)$, where N is the number of spatial dimensions. $\lambda$ is the eigenvalue while $Q$ is the linearization operator (usually non-Hermitian) and  is given by
 \begin{widetext}
  \begin{eqnarray}
 Q=\frac{1}{2}
 \left(
 \begin{array}{cc}
 (1-\eta_2\partial_y^2)^{-1}(G_1\partial_y-\eta_1\partial^3_y-\nu\partial_y^2 +G_0) \hskip 0.5truecm & \hskip 0.5truecm (\eta_2\partial_y^2-1)^{-1}G_2 \\
 (\eta_2\partial_y^2-1)^{-1}G_2\hskip 0.5truecm & (1-\eta_2\partial_y^2)^{-1}(G_1\partial_y-\eta_1\partial^3_y-\nu\partial_y^2 +G_0) \hskip 0.5truecm \\
 \end{array}
 \right) 
 \end{eqnarray}
 \end{widetext}
 \begin{eqnarray}
 \begin{cases}
 G_0=-2\beta A^2\textrm{ sech}^2(ay) \textrm{tanh}(ay), \\
 G_1=c_0^2+2\beta A^2 \textrm{ sech}^2(ay), \\
  G_2= 2aA^2\textrm{ sech}^2(ay)\textrm{ tanh}(ay).
 \end{cases}
\end{eqnarray}
The continuous spectrum of the system is determined analytically by examining the matrix $Q$ as $\textbf{y} \rightarrow \infty$. Proceeding with  this approximation, it is easy to show that the continuous eigenvalues are
\begin{eqnarray}
\lambda=\pm \frac{1}{2}\left \lbrace  i\frac{c_0^2k+\eta_1k^3}{1+\eta_2k^2}+\frac{\nu k^2}{1+\eta_2k^2}\right \rbrace.
\end{eqnarray}
   where $k$ is the perturbation wave number. Discrete eigenvalues do exist, however, they can only be obtained numerically. The local growth rate of the solitons or the gain is  given by $Im(\lambda)$. The maximum growth rate of the soliton  can be obtained by solving the equation $\frac{d Im(\lambda)}{dk}=0$. Applying this definition, it is easy to show that,
   \begin{eqnarray}
    k^2_{\pm}=\frac{\eta_2c_0^2-3\eta_1 \pm \sqrt {(3\eta_1-\eta_2c_0^2)^{2}-4\eta_1\eta_2c_0^2 }}{2\eta_1\eta_2}.\qquad
   \end{eqnarray}
  \begin{figure}[ht]
  \includegraphics[width=3.5in, height= 3.2in]{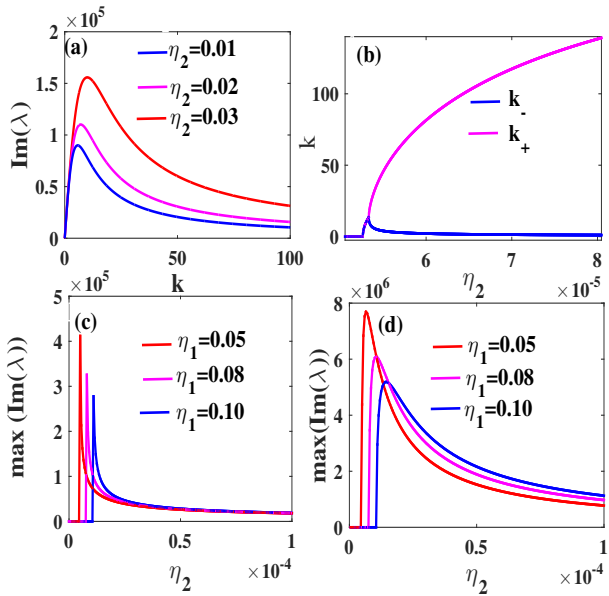}
  \caption{\label{fig9}(a)The perturbation growth rate of solitons versus $k$ with different values of $\eta_2$, $\eta_1=0.05$ (b) Bifurcation diagram of  soliton states in the two components model. $0<\eta_2<5.32 \times 10^{-5} $, the wave number $k=0$,  $k=5.1\times 10^{-5}$,  there is a growth in the wave number $k$ and at $k=5.32 \times 10^{-5}$, the soliton split into two soliton states with wave number $k_{-}$ and $k_+$. Note that their antisymmetric components $-k_{-}$ and $-k_{+}$ are not shown on the diagram and $\eta_1=0.55$ (c) and (d)  corresponds to the maximum perturbation growth rate of solitons states $k_{-}$ and  $k_{+}$  respectively versus $\eta_2$ with different values of $\eta_1$. The order parameter  $c_0=176.6$. }
  \end{figure}

The stability of the perturbed soliton $u, v$ is related to the imaginary parts
$Im(\lambda)$ of all eigenvalues $\lambda$. If $|Im(\lambda)| >0$, then the solution
$(u, v)$ will grow exponentially with $y$ (i.e., it is unstable),
otherwise, the solution $(u, v)$ is stable. In the existence
range of the approximation described above, the dependence of
the perturbation growth rate (the most unstable growth rate) on the parameter $\eta_1$ and $\eta_2$ are exhibited Fig. \ref{fig9} (a) and 5(c-d) respectively. As it can clearly be observed, $\eta_1$ and $\eta_2$ reduces the instability growth rate of the solitons.\

\begin{figure}[ht]
  \includegraphics[width=3.68in, height= 3.58in]{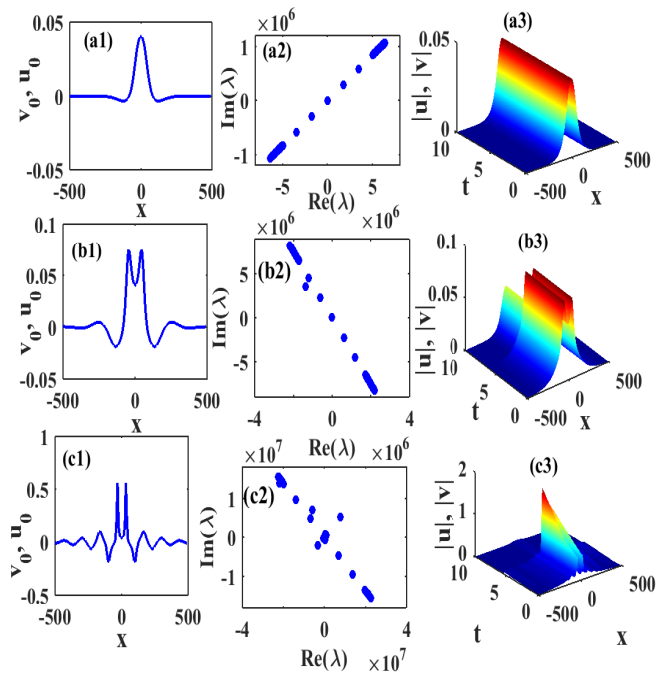}
  \caption{\label{figa2} (a1, b1, c1) Initial solitary wave profile as given by  Eq. (\ref{e9}). (a2, b2, c2) Numerically computed linear stability spectra. (a3) Stable and (b2, c3) unstable propagations of perturbed solution $u=u_0+\varphi u_1$ and $v=v_0+\varphi v_1$. (a1,
  a2, a3) $\xi=165.8$, (b1, b2, b3) $\xi=165.5$, and (c1, c2, c3) $\xi=165.0$.  The other model parameters
    are $\epsilon=0.001$,  $\eta_1=0.045$, $\nu=0.05$, 
     $\rho^A_0=4.035 \times 10^{-3}$, $\varphi=10^{-33}$, $\eta_2=0.11$, and $\beta= 79.5 \frac{c_0^2}{(\rho^A_0)^2}$. }
  \end{figure}

 In order to confirm the stability or instability of the system, one needs to solve the eigenvalue equation (\ref{f3}) numerically. One of the methods used to solved such a problem is the finite difference discretizations method. However, the accuracy of this method is quite low. In addition, this method has been reported to give spurious eigenfunction even when the eigenvalues obtain are approximately correct. A more accurate method is the Fourier collocation method (FCM) \cite{F20}. By this method, the eigenfunction $\textbf{V}$ is expanded into a Fourier series
and Eq. (\ref{f3}) is turned into a matrix eigenvalue problem for the Fourier coefficients of the
eigenfunction \textbf{V}. The structure of the resulting matrix depends heavily on the structure of the spatial dimension of  Eqs. (\ref{e4a}) and (\ref{e4b}). A detail analysis of the Fourier collection method can be found in \cite{F20, F21, F22}. The results from the numerical simulation is illustrated in Fig.  ( \ref{figa2}).

  Figure \ref{figa2} (a1, b1, c1) illustrate the unperturbed soliton profiles for different values of the propagation speed $\xi$ used in our numerical simulation while Fig. \ref{figa2}  (a2, b2, c3)  corresponds to the numerically computed eigenvalue spectrum. As can clearly seen, Fig. \ref{figa2} (a2)  describe a bound state or internal mode. In particular, persistent oscillations of the soliton
width, and position are due to these modes. On the other hand, Fig. \ref{figa2} (a3) illustrate to the stable propagation of the perturbed solution while Fig. \ref{figa2} (b3, c3) illustrate unstable propagation of the intensity of the perturb solution.

\newpage
\section{Discussion and Conclusion}
Electromechanical solitary wave measurements on
nerve fibers began in the early 1980s with the measuring
of various non-electrical components of action potential
and the investigation of the chemistry of phase transitions in nerve fibers, and its importance for nerve
pulse propagation by Tasaki \cite{h1} and by Kaufmann proposal
of sound waves as a physical basis for action potential
propagation in the nerve. This mechanical aspect of
action potential gained more attention when Heimburg
and Jackson proposed the soliton model for nerve pulse
propagation in the early 2000s \cite{F4}. Since that time, the dynamics
of a pulse, propagation has been extensively studied by a
a wide variety of applied scientists; and, partly as a result
of recent interest shown by applied mathematicians
in the nonlinear diffusion equation, single pulse dynamics is
now a rather well understood physical phenomenon \cite{F8, F9, F10, s6, s7, F11, F12}. One might suppose that this is the end of the story;
on the contrary, it is only the beginning. Recent studied
have shown that the electromechanical density pulses
in the nerve can effectively be considered as a vector
soliton with two components. That is, the longitudinal
component corresponding relative height field and the
transverse components corresponding to the lateral stretch
field and have been observed experimentally in garfish
olfactory nerve, squid giant axon, and hippocampal
neuron \cite{F13}.\

 In this work, we have considered the
electromechanical density pulse as a two coupled solitary
waves represented by longitudinal compression wave
and an out-of-plane transversal pulse (i.e., perpendicular
to the membrane surface) and analyzed using the variational
approach the characteristics of the coupled solitary
waves in the presence of damping within the framework of
coupled nonlinear Burger-Korteweg-de Vries-Benjamin-
Bona-Mahony equations(BKdV-BBM) derived from the
vector solito model equation for biomembranes and nerves.
In particular, we have shown that, there must be a balanced between damping and inertia effects for a stable coupled solitary waves to propagates within the axon. Furthermore, we have shown that the presence of damping coefficient causes a discontinuity in the ($\eta_2$, $\xi$) parameter space. In particular, we observed a discontinuity in stable regions stable region which appears like the island of points. Analysis of the solitary wave energy shows that both $\eta_2$ and $\nu$ causes a decrease the energy of the coupled solitary waves. In addition,  the potential function of interaction is shown to be always negative. Thus, the interaction between the two solitary waves is attractive; which is agreement with the fact of neural coupling oscillation. We performed linear stability analysis and the results shows that both $\eta_2$ and $\nu$ decreases the perturbation growth rate of the coupled solitary waves.
Numerical simulation of the linearized equation shows
that the bell shape solitary wave profile yields a bound
eigenvalue spectrum while the solitary shock-like solitary
waves produce an unbound eigenvalue spectrum. Thus
the bell shape profile is generally stable for a small perturbation
while the solitary-shock like profile is generally unstable.
The instability of these shock-like solitary waves
might be responsible for traumatic brain injury and damage
to the cell membrane \cite{SS1}.

\end{document}